\documentclass[aps,a4paper,twocolumn,superscriptaddress]{revtex4}
\usepackage{graphicx}

\newcommand{\beq}{\begin{equation}}
\newcommand{\eeq}{\end{equation}}
\newcommand{\beqa}{\begin{eqnarray}}
\newcommand{\eeqa}{\end{eqnarray}}

\renewcommand{\a}{\alpha}
\newcommand{\abs}[1]{\vert#1\vert}
\renewcommand{\b}{\beta}
\renewcommand{\d}{{\rm d}}
\newcommand{\del}{\delta}
\newcommand{\e}{{\rm e}}
\newcommand{\eff}{{\rm eff}}
\newcommand{\eps}{\varepsilon}
\newcommand{\g}{\gamma}
\newcommand{\frad}[2]{{\displaystyle{#1\over#2}}}
\newcommand{\frat}[2]{{\textstyle{#1\over#2}}}
\newcommand{\loc}{{\rm loc}}
\renewcommand{\o}{\omega}
\newcommand{\s}{\sigma}
\renewcommand{\L}{{({\rm L})}}
\renewcommand{\O}{\Omega}
\newcommand{\R}{{({\rm R})}}

\begin{document}

\title{Slow synaptic dynamics in a network: from exponential to power-law
forgetting}

\author{J. M. Luck}\email{jean-marc.luck@cea.fr}
\affiliation{Institut de Physique Th\'eorique, URA 2306 of CNRS, CEA Saclay,
91191 Gif-sur-Yvette cedex, France}

\author{A. Mehta}\email{anita@bose.res.in}
\affiliation{S. N. Bose National Centre for Basic Sciences, Block JD,
Sector 3, Salt Lake, Calcutta 700098, India}

\begin{abstract}
We investigate a mean-field model of interacting synapses on a directed neural network.
Our interest lies in the slow adaptive dynamics of synapses,
which are driven by the fast dynamics of the neurons they connect.
Cooperation is modelled from the usual Hebbian perspective,
while competition is modelled by an original polarity-driven rule.
The emergence of a critical manifold culminating in a tricritical point
is crucially dependent on the presence of synaptic competition.
This leads to a universal $1/t$ power-law relaxation of the mean synaptic
strength along the critical manifold
and an equally universal $1/\sqrt{t}$ relaxation at the tricritical point,
to be contrasted with the exponential relaxation that is otherwise generic.
In turn, this leads to the natural emergence of long- and short-term memory
from different parts of parameter space in a synaptic network, which is the
most novel and important result of our present investigations.

\end{abstract}

\maketitle

\section{Introduction}
\label{intro}

Memory and its mechanisms have always attracted a great deal of interest~\cite{ebb}.
It is well known that memory is not a monolithic
construct, and that memory subsystems corresponding to episodic, semantic or
working memory exist~\cite{squire}.
We focus here on explicit memory, which is the memory for events and facts.

In general, memories are acquired by the process of
learning, which is a complicated phenomenon related to neural activities,
brain network structure and synaptic plasticity~\cite{review}.
However, neuroscientists~\cite{turr2,avy} typically focus on the latter, so that
increasingly sophisticated models of synaptic plasticity
have emerged over the years~\cite{book1,book2,book3}.
Much of this work has been done by adapting methods from statistical physics.
Such modelling, while it may not include details of specificities involving chemical and
biological processes in the brain, can outline possible mechanisms that take
place in simplified structures.
For example, the study of neural networks~\cite{book1,book2,book3}, while it greatly
simplifies biological structures in order to make them tractable,
has still been able to make an impact on the parent field.
In particular, neural networks such as the Hopfield model~\cite{hop1,hop2}
have been extensively investigated via methods borrowed from the statistical physics
of disordered and complex systems~\cite{ags1,ags2,ags3,ags4}.
In these models, memories are stored as patterns of neural activities,
which correspond both to low-energy states
and to attractors of the stochastic dynamics of the model.
An essential property of these models as well as real neural networks is that
their capacity is finite.
Forgetting is therefore an important aspect of continued
learning~\cite{nadal,parisi,landf1,landf2,landf3,landf4,review}.

More recently, there has been a great deal of work on the {\it fast} dynamics
of neurons in neural networks.
Typically, models of integrate-and-fire
neurons on networks have been studied, and their different dynamical regimes
explored~\cite{brunel}.
The discovery of neural avalanches
in the brain~\cite{beggs} was followed by several dynamical models of
neural networks~\cite{born,roxin}, where the statistics of avalanches were
investigated~\cite{lucilla1,lucilla2,geisel1,geisel2,geisel3,taylor} in the
context of theories of self-organised criticality~\cite{bak}.
A review of such approaches can be found in~\cite{rabi}.

Here, by contrast, we study the {\it slow} dynamics of adaptive
synapses in neural networks.
This is done with the objective of exploring
the phenomena of learning and forgetting, to both of which the evolution of
synaptic plasticity is strongly linked~\cite{review}.

We usually tend to remember information only for relatively short durations:
such finite time scales, corresponding to short-term memory, are readily
modelled by a process of exponential forgetting.
However, there are some things
we remember for as long as we live, which form part of our long-term memories:
this scenario corresponds to {\it power-law forgetting}~\cite{plf1,plf2,plf3},
with its attendant absence of time scales.
Typically, models which
manifest the latter have made use of specially designed synapses with
`hidden' internal states~\cite{fusi,jstat,plosone}.
The aim of this paper is to provide
a holistic framework for the modelling of synaptic networks which are {\it capable of
storing both long- and short-term memories, without recourse to specialised architectures}.
In our model, these emerge naturally in different parameter regimes,
as a direct consequence of the collective dynamics of synaptic cooperation and competition.
Our model thus provides a clear modelling
alternative to the cascade process of Fusi and collaborators~\cite{fusi} which have so far
occupied centre stage in the field: while their model invokes specialised synaptic architectures
to get long-term memory, ours does not.

In general, neural processes are assumed to be subject to local rules that
govern the way in which synapses are updated~\cite{barrett,review}; Hebb's rule
is an important example which
says that `neurons that fire together, wire together'~\cite{hebb}.
The outcome of many such processes results in functional change which drives
behaviour, in much the same way as in agent-based modelling, when local interactions
among agents may give rise to emergent phenomena on a macroscopic scale~\cite{santo}.
In such approaches~\cite{uspr}, the underlying idea is that the strategy of a
given agent is to a large extent determined by what the others
are doing, through considerations of the relative payoffs obtainable in each case.
This formalism was extended to neural networks in
a simple-minded way in~\cite{gmam1,gmam2}, where synapses adapted via {\it competing}
interactions involving the activity patterns of interconnected neurons.

This paper puts that earlier work on a more complete footing, in particular by
extending the types of synaptic interactions.
It is known that both competition
and cooperation play important roles in synaptic plasticity~\cite{miller}; cooperation
has traditionally been modelled by Hebb's rule, but this alone can lead to the
unlimited growth of synaptic strength, which is unphysical.
Competition is thus a
necessary mechanism to regulate such growth~\cite{turr1,turr2}: while regarded
as an essential ingredient by neuroscientists, its inclusion in theoretical models
is rare~\cite{avy}.
An example of competition in the fast dynamical regime
of firing neurons can be found in~\cite{song}, where synaptic updates occur
depending on the latency of spike trains.
Our modelling of competition~\cite{gmam1} is, however,
formulated in the opposite dynamical regime of slow synaptic dynamics.
Finally, we also include a representation of the spontaneous relaxation of synapses.
This mechanism is an important one in the context of finite storage capacities,
when space is created via the spontaneous decay of old memories.
This is sometimes referred to as the palimpsest effect~\cite{nadal,parisi}.

At the most microscopic level, individual neurons fire at rates that
exhibit a whole spectrum of biological noise~\cite{book1,book2,book3}.
Here we choose a level of description where neurons may be either active or inactive,
according to their mean firing rates.
The response of neurons is considered as stochastic
and instantaneous with respect to the much slower dynamics of the synapses we consider.
As a result of this temporal coarse-graining,
the overall effect of the microscopic noise can be represented
by spontaneous relaxation rates from one type of synaptic strength to the other.
Next, cooperation between synapses is incorporated via the usual Hebbian viewpoint.
Our modelling of the competitive interaction by polarity-driven interactions is
the most original as well as the most crucial part of our formalism: synapses
are converted to the type most responsible for neural activity in their
neighbourhood~\cite{gmam1,gmam2}.

Our choice of basis is that of a fully connected network,
where all neurons are connected to one another by directed synapses~\cite{book1,book3}.
Section~\ref{model} contains a detailed description of the model.
In section~\ref{mf}, we characterise the various types of mean-field
dynamics (generic, critical, tricritical) displayed by our model.
In Section~\ref{diagram}, we explore the dependence of our phase diagram on parameters,
with particular reference to the behaviour of relaxation times.
In Section~\ref{landf}, we address issues related to learning and forgetting,
and show that our model contains a rich spectrum of time scales.
Finally, in Section~\ref{disc}, we discuss our findings.

\section{The model}
\label{model}

We model a network of neurons connected by directed synapses.
We first describe the geometry of our network and then explain the nature of its dynamics.

\subsection{Geometry}

\begin{figure}[!ht]
\begin{center}
\includegraphics[angle=-90,width=.55\linewidth]{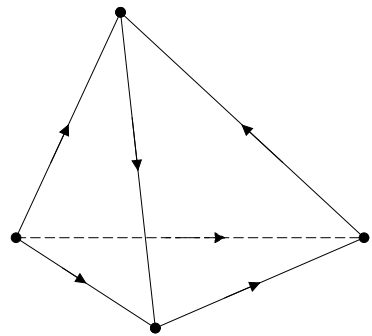}
\caption{\small
An instance of the directed fully connected network with 4 nodes.}
\label{dirmeanfield}
\end{center}
\end{figure}

We consider a fully connected network,
whose bonds are directed (see Figure~\ref{dirmeanfield})
by randomly attributing an orientation to every bond
of the complete graph on $N$ nodes.
With this geometry, mean-field theory is expected to apply
in the thermodynamic limit of an infinitely large network~\cite{book1}.

We mention in passing that our assumption of a fully connected network in our analysis
is only for technical simplicity. In fact, mean-field dynamics 
are also expected to apply to sparse networks,
provided that 
the degree (number of neighbours) of its nodes grows to infinity
with the system size $N$. Realistic neural networks, while sparse,
indeed show such growth~\cite{STN,SSR}, so that it is valid to do a mean-field
analysis of their dynamics. It turns out that for such networks,
the degree grows linearly with system size; it is of the form $pN$
where $p$ is of the order of 10 to 15 percent.

Neurons live on the nodes of the network, labelled $i=1,\dots,N$.
The activity state of neuron $i$ at time $t$ is described
by a binary activity variable:
\beq
\nu_i(t)=\left\{\matrix{
+1&\mbox{if $i$ is {\it active}{\hskip 9pt} at time $t$},\hfill\cr
-1&\mbox{if $i$ is {\it inactive} at time $t$}.\hfill\cr
}\right.
\eeq
Active neurons are those whose instantaneous firing rate
exceeds some unspecified threshold.

The network is equipped with oriented bonds as follows.
For each pair of different nodes $i$ and $j$,
we attribute a random orientation to the bond joining $i$ and $j$,
i.e., we put with probability $\frat12$ either a directed bond $(ij)$ from $i$ to $j$,
or a directed bond $(ji)$ from $j$ to $i$, but never both.
The total number of oriented bonds is therefore $\frat12N(N-1)$.
Synapses live on the directed bonds $(ij)$ so defined.
The strength $\s_{ij}(t)$ of synapse $(ij)$ at time $t$
is also described by a binary variable:
\beq
\s_{ij}(t)=\left\{\matrix{
+1&\mbox{if $(ij)$ is {\it strong} at time $t$},\hfill\cr
-1&\mbox{if $(ij)$ is {\it weak}{\hskip 5pt} at time $t$}.\hfill\cr
}\right.
\eeq
Strong synapses are those whose strength exceeds some unspecified threshold.

\subsection{Neuronal dynamics}

Neurons have an instantaneous stochastic response to their environment.
The activity of neuron $i$ at time $t$ reads
\beq
\nu_i(t)=\left\{\matrix{
+1&\mbox{with probability $F(h_i(t))$},\hfill\cr
-1&\mbox{with probability $1-F(h_i(t))$},\hfill\cr
}\right.
\label{nudef}
\eeq
where $F(h)$ is a sigmoidal response function of the input field $h_i(t)$,
increasing from $F(-\infty)=0$ to $F(+\infty)=1$.
The input field acting on neuron $i$,
\beq
h_i(t)=\frac{1}{N}\sum_{j\in\partial(i)}(a+b\s_{ji}(t))\nu_j(t),
\label{hdef}
\eeq
is a weighted sum of the instantaneous activities $\nu_j(t)$
of the neurons $j$ which influence $i$.
Here, $\partial(i)$ denotes the subset of nodes $j$ which transmit information to $i$
via directed synapses $(ji)$.
Strong synapses ($\s_{ji}=1$) enter the above sum with a synaptic weight $a+b$,
while weak ones ($\s_{ji}=-1$) have a synaptic weight $a-b$.
We assume $a$ and $b$ are constant all over the network.

All synapses are therefore excitatory for $b>0$, and inhibitory for $b<0$.
The kind of collective behavior discussed here,
either along the critical manifold of at its tricritical endpoint,
is therefore different in nature from the chaotic dynamical features which have been
emphasized in balanced networks~\cite{VS1,VS2,BCD},
where excitatory and inhibitory effects balance each other on average.

In the following, we consider a spatially homogeneous situation
in the thermodynamic limit of an infinitely large network.
In this limit, for every node $i$,
the numbers of incoming bonds $(ji)$ and outgoing bonds $(ik)$
are both equal to $\frat12N$, up to negligible fluctuations.
Moreover, we focus on the slow plasticity dynamics of the synaptic strengths.
The characteristic time scale of this dynamics is much larger than
the microscopic time scale of neural activity.
Within this framework, it will be sufficient to consider the mean neural activity
\beq
A(t)=\frac{1}{N}\sum_i\nu_i(t)
\eeq
and the mean synaptic strength
\beq
J(t)=\frac{2}{N(N-1)}\sum_{(ij)}\s_{ij}(t).
\label{jdef}
\eeq
These key dynamical quantities entirely characterise
the global aspects of the slow synaptic dynamics.
They are related by a {\it constitutive equation} of the form
\beq
A(t)=g(J(t)),
\label{aj}
\eeq
which is local in time:
the mean neural activity $A(t)$ only depends on the mean synaptic strength $J(t)$
at the same time $t$.
The form of the function $g(J)$ can, at least in principle,
be derived by appropriately averaging
the microscopic equations~(\ref{nudef}) and~(\ref{hdef}),
both spatially over the network and temporally over an integration time $\Delta t$,
which would be large with respect to the time interval between two spikes, say,
and very small with respect to the characteristic time scale of plasticity dynamics.

In this work, we prefer to employ a more phenomenological route.
Remember that all the synapses of the network are excitatory for $b>0$,
and inhibitory for $b<0$.
Consider for a while the special situation where there are as many strong as weak synapses.
In this case the mean synaptic strength defined in~(\ref{jdef}) vanishes ($J=0$).
We make the simplifying assumption that there are also as many active as inactive
neurons on average in this situation, so that $A=0$.
Then, linearising the constitutive equation~(\ref{aj}) around this symmetric situation,
we readily obtain the linear response formula
\beq
g(J)=\eps J,
\label{eff}
\eeq
which will be used throughout this work.
The slope $\eps$ of the response function is one of the key parameters of the model.
It is clearly proportional to $b$, and positive in the excitatory case ($b>0$),
so that $g(J)$ is an increasing function of $J$.
In the inhibitory case ($b<0$), $\eps$ is negative,
so that $g(J)$ is a decreasing function of $J$.
Finally, it has to obey $\abs{\eps}<1$.

\subsection{Synaptic plasticity dynamics}

Synaptic strengths evolve very slowly in time,
compared to the fast time scales of neuron firing rates.
It is thus natural to model synaptic dynamics as a stochastic process
in continuous time~\cite{kampen},
defined in terms of effective jump rates between the two states,
strong or weak, of the synaptic strength.
Our model includes three plasticity mechanisms which drive synaptic evolution:

\subsubsection{Spontaneous relaxation mechanism}

Synapses may spontaneously change their state
from weak to strong (potentiation) or strong to weak (depression).
This spontaneous relaxation mechanism translates into
\beq
\left\{\matrix{
\s_{ij}=-1\to+1&\mbox{with rate $\O$},\hfill\cr
\s_{ij}=+1\to-1&\mbox{with rate $\o$}.\hfill\cr
}\right.
\label{m1}
\eeq
Signal processing, in this context,
examines the effect of deterministic external signals which are superposed
on these spontaneous relaxation rates (see~(\ref{local})).

\subsubsection{Hebbian mechanism}

When two neurons are in the same state of (in)activity,
the synapse which connects them strengthens;
when one of the neurons is active and the other is not,
the interconnecting synapse weakens.
This is the well-known Hebbian mechanism~\cite{hebb},
which we implement with rate $\a$.
In the thermodynamic limit of the directed fully connected network,
the probability to have $\nu_i=\pm1$ at any given time
is $\frat12(1\pm A)=\frat12(1\pm g(J))$.
The probabilities $q_+$ (resp.~$q_-$)
to have $\nu_i=\nu_j$ (resp.~$\nu_i\ne\nu_j$) are:
\beq
q_\pm=\frat12(1\pm g(J)^2).
\eeq
We thus have
\beq
\left\{\matrix{
\s_{ij}=-1\to+1&\mbox{with rate $\a q_+$},\hfill\cr
\s_{ij}=+1\to-1&\mbox{with rate $\a q_-$}.\hfill\cr
}\right.
\label{m2}
\eeq

\subsubsection{Polarity-driven mechanism}

This is a mechanism to introduce synaptic competition,
introduced for the first time in~\cite{gmam1},
which converts a given synapse to the type of its most `successful' neighbours.
Thus: if a synapse $(ij)$ connects two neurons with different
activities at any given time,
it will adapt its strength to that of a randomly selected synapse
connected to the active neuron.
If we have $\nu_i=+1$ and $\nu_j=-1$,
the active neuron $i$ is {\it presynaptic},
and the selected synapse may be either outgoing $(ik)$ from neuron~$i$,
or incoming $(ki)$ to neuron $i$.
If we have $\nu_i=-1$ and $\nu_j=+1$,
the active neuron $j$ is {\it postsynaptic},
and the selected synapse may be either outgoing $(jk)$ from neuron $j$,
or incoming $(kj)$ to neuron $j$.
If the selected synapse is strong,
the update $\s_{ij}=-1\to+1$ takes place with rate $\b$;
if it is weak,
the update $\s_{ij}=+1\to-1$ takes place with rate $\g$.
The rates $\b$ and $\g$ are assumed to be identical in all four cases.

All in all, the polarity-driven mechanism also translates into a simple form
in the thermodynamic limit of the directed fully connected network:
\beq
\left\{\matrix{
\s_{ij}=-1\to+1&\mbox{with rate $\frat12\b(1+J)q_-$},\hfill\cr
\s_{ij}=+1\to-1&\mbox{with rate $\frat12\g(1-J)q_-$}.\hfill\cr
}\right.
\label{m3}
\eeq

\section{Mean-field dynamics}
\label{mf}

Here we begin our investigation of the slow collective dynamics of
the synaptic activity in the network:
importantly, we restrict ourselves to its global features,
rather than looking at patterns of spatially varying synaptic strengths.

For a spatially homogeneous situation in the thermodynamic limit,
the mean synaptic strength $J(t)$
obeys a non-linear dynamical mean-field equation of the form
\beq
\frac{\d J}{\d t}=P(J).
\label{djdt}
\eeq
The explicit form of the rate function $P(J)$ is obtained by summing the
contributions of the three plasticity mechanism mentioned above:
\beq
P(J)=P_1(J)+P_2(J)+P_3(J),
\eeq
with (see~(\ref{m1}), (\ref{m2}), (\ref{m3}))
\beqa
P_1(J)\!\!&=&\!\!\O(1-J)-\o(1+J),
\nonumber\\
P_2(J)\!\!&=&\!\!\a\left(g(J)^2-J\right)
\nonumber\\
\!\!&=&\!\!-\a J(1-\eps^2J),
\nonumber\\
P_3(J)\!\!&=&\!\!-\del(1-J^2)(1-g(J)^2)
\nonumber\\
\!\!&=&\!\!-\del(1-J^2)(1-\eps^2J^2),
\label{p123}
\eeqa
where
\beq
\del=\frat{1}{4}(\g-\b).
\eeq

In the most general situation, the model has five parameters:
the slope $\eps$ of the linear response equation~(\ref{eff})
and the four rates $\O$, $\o$, $\a$, and $\del$
involved in the three plasticity mechanisms.
The resulting rate function is a polynomial of degree 4:
\beq
P(J)=p_4J^4+p_2J^2-(\O+\o+\a)J+\O-\o-\del,
\label{pj}
\eeq
with
\beq
p_4=-\del\eps^2,\quad p_2=(\a+\del)\eps^2+\del.
\label{p2p4}
\eeq
The linear rate function $P_1(J)$ corresponds to the spontaneous mechanism;
the Hebbian mechanism leads to the quadratic non-linearity of $P_2(J)$,
while the polarity-driven competitive mechanism results in the quartic
non-linearity of $P_3(J)$.

The parameter $\eps$ only enters~(\ref{p123}) and~(\ref{p2p4}) through its square $\eps^2$.
The model therefore exhibits an exact symmetry
between the excitatory ($\eps>0$) and the inhibitory ($\eps<0$) cases.
This is to be expected, as none of the plasticity mechanisms distinguishes between them.
More generally,
the model is invariant if the constitutive function $g(J)$ is changed into its opposite.

\subsection{Generic dynamics}

\begin{figure}[!ht]
\begin{center}
\includegraphics[angle=-90,width=.65\linewidth]{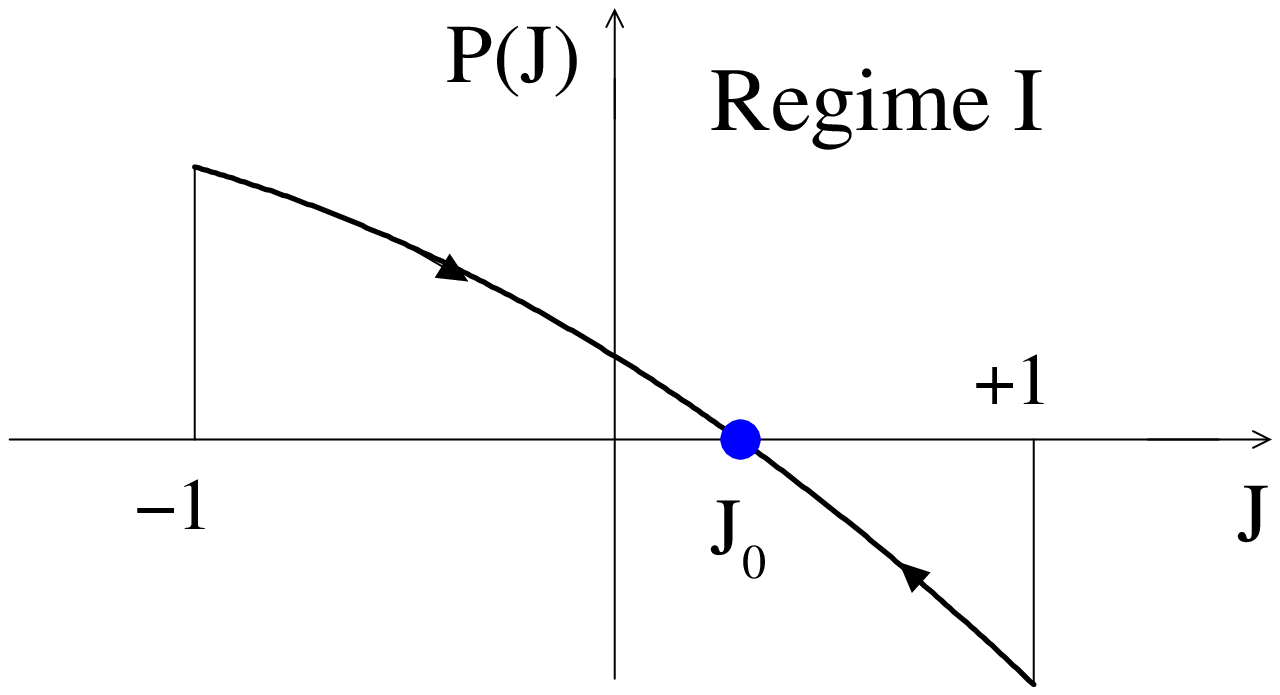}

\includegraphics[angle=-90,width=.65\linewidth]{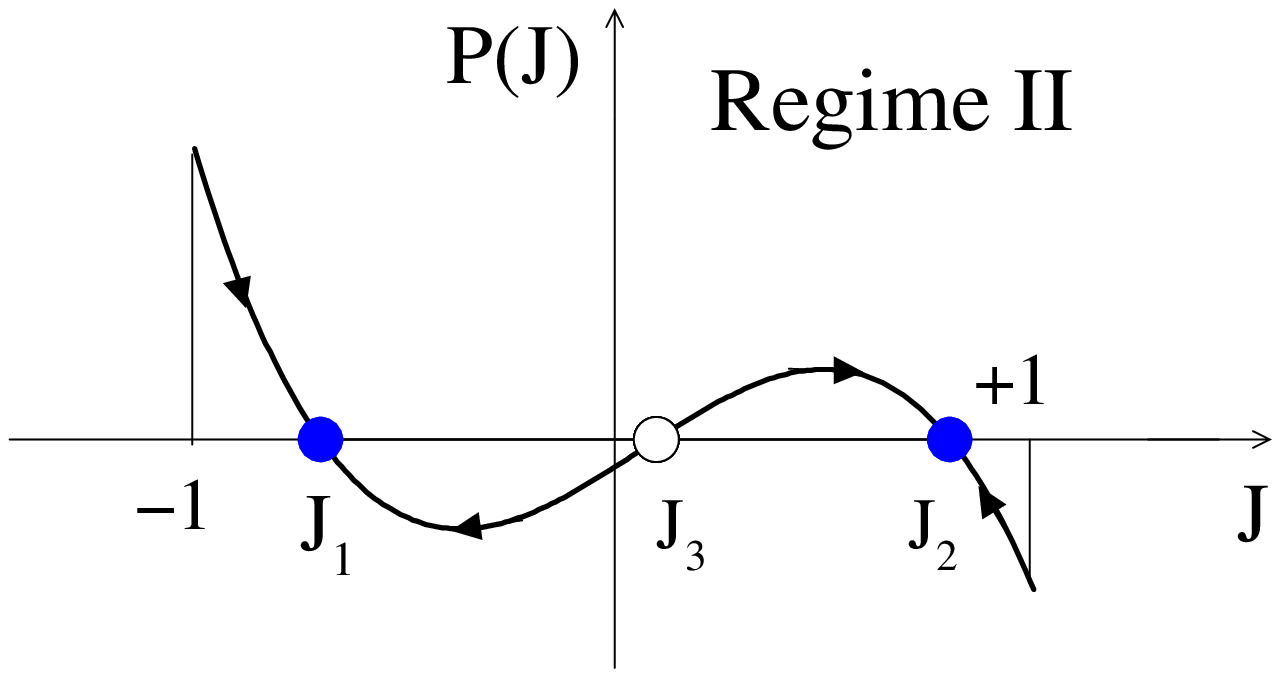}
\caption{\small
(Color online)
The two possible generic dynamical regimes.
Top: Regime~I (one single attractive fixed point, $J_0$).
Bottom: Regime~II (two attractive fixed points, $J_1$ and $J_2$,
and an intermediate repulsive one, $J_3$).}
\label{regimes}
\end{center}
\end{figure}

The dynamics leave the mean synaptic strength
confined to the physical interval $-1\le J(t)\le 1$.
We have indeed $P(-1)=2\O+\a(1+\eps^2)>0$ and $P(1)=-2\o-\a(1-\eps^2)<0$.
The rate function $P(J)$ has therefore an odd number of zeros in this interval,
i.e., either one or three, with appropriate multiplicities in critical regimes.
These zeros correspond to fixed points of the dynamics.
As a consequence, the model exhibits two generic dynamical regimes,
as shown in Figure~\ref{regimes}.

In Regime~I (see Figure~\ref{regimes}, top),
there is a single attractive (stable) fixed point at $J_0$.
The mean synaptic strength $J(t)$ therefore converges exponentially fast
to this unique fixed point, irrespective of its initial value, according to
\beq
J(t)-J_0\sim\e^{-t/\tau_0}.
\eeq
The corresponding relaxation time reads
\beq
\tau_0=-\frac{1}{P'(J_0)}.
\label{tau0}
\eeq
In the limiting situation where there is only spontaneous relaxation,
so that $P(J)=P_1(J)$, we have
\beq
J_0=\frac{\O-\o}{\O+\o},\quad\tau_0=\frac{1}{\O+\o}.
\eeq

In Regime~II (see Figure~\ref{regimes}, bottom),
there are two attractive (stable) fixed points at $J_1$ and $J_2$,
and an intermediate repulsive (unstable) one at $J_3$.
The mean synaptic strength $J(t)$ converges exponentially fast to
either of the attractive fixed points, depending on its initial value,
namely to $J_1$ if $-1<J(0)<J_3$ and to $J_2$ if $J_3<J(0)<1$.
The corresponding relaxation times read
\beq
\tau_1=-\frac{1}{P'(J_1)},\quad\tau_2=-\frac{1}{P'(J_2)}.
\label{tau12}
\eeq
In other words, Regime~II allows for the coexistence of two separate fixed points,
leading to network configurations which are composed of largely strong/weak synapses.
The quartic non-linearity corresponding to
the polarity-driven plasticity mechanism needs to be sufficiently strong for
the model to exhibit this coexistence (see Section~\ref{diagram}).

\subsection{Critical dynamics}

\begin{figure}[!ht]
\begin{center}
\includegraphics[angle=-90,width=.65\linewidth]{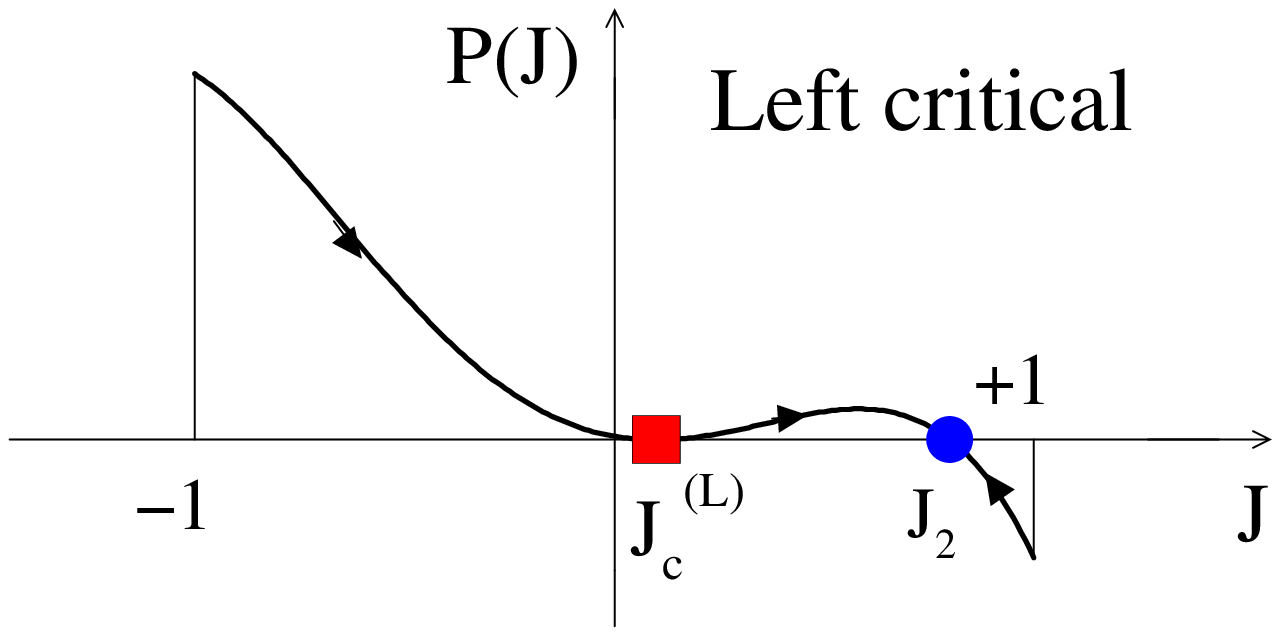}

\includegraphics[angle=-90,width=.65\linewidth]{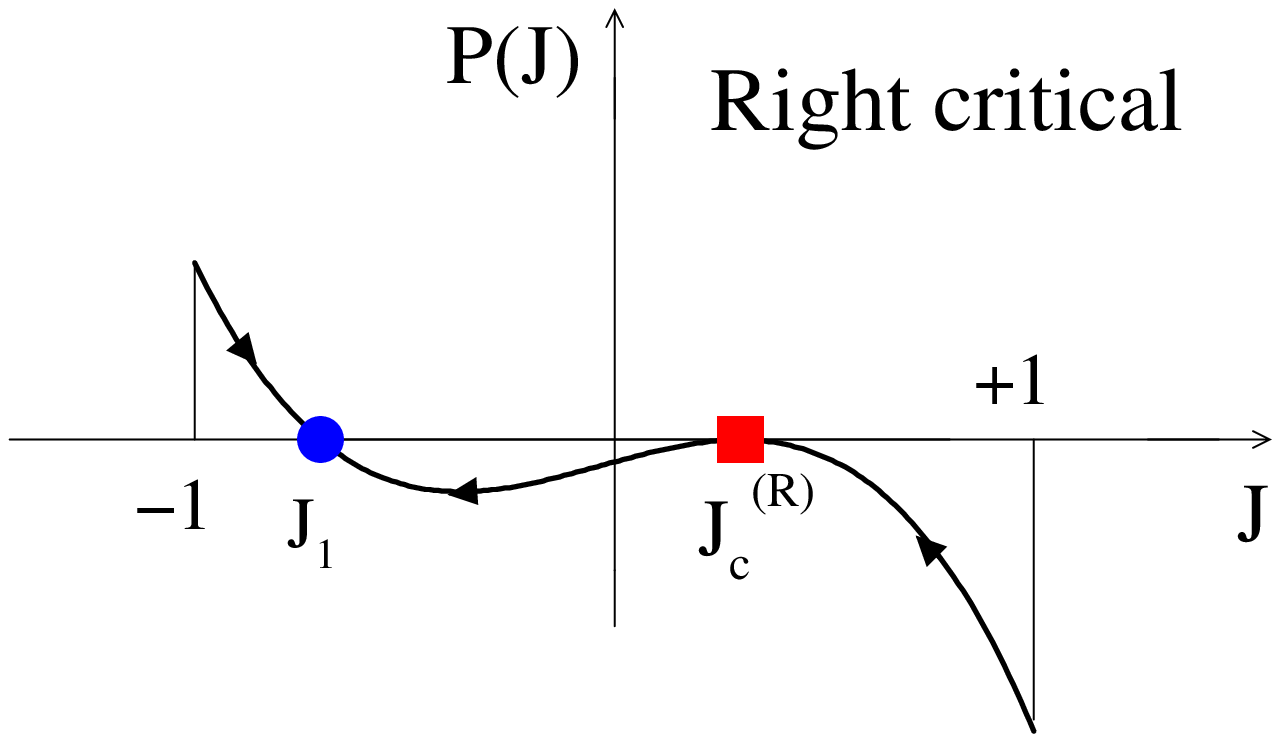}
\caption{\small
(Color online)
The two possible kinds of critical dynamical behaviour.
Top: left critical case ($J_1=J_3=J_c^\L$).
Bottom: right critical case ($J_2=J_3=J_c^\R$).}
\label{critical}
\end{center}
\end{figure}

When two of the three fixed points merge at some $J_c$,
the dynamical system~(\ref{djdt}) exhibits a saddle-node bifurcation~\cite{bif}.
In physical terms, the dynamics become critical.
We have then
\beq
P(J_c)=P'(J_c)=0,
\label{double}
\eeq
so that the critical synaptic strength $J_c$
is a double zero of the rate function $P(J)$ (see Figure~\ref{critical}).
There is a left critical case, where $J_1=J_3=J_c^\L$, while $J_2$ remains non-critical,
and a right one, where $J_2=J_3=J_c^\R$, while~$J_1$ remains non-critical.
The critical synaptic strength $J_c$ in both cases
will be shown to obey $J_c>\frat{1}{3}$ (see~(\ref{tiers})).
We thus conclude that the critical point is always {\it strengthening},
as $J_c$ is always larger then the `natural' initial value $J(0)=0$,
corresponding to a random mixture of strong and weak synapses in equal proportions.

The mean synaptic strength exhibits a universal relaxation
to its critical value, of the form
\beq
J(t)-J_c\approx\frac{A_c}{t}.
\label{ctime}
\eeq
The corresponding amplitude reads
\beq
A_c=-\frac{2}{P''(J_c)}=\frac{1}{6\del\eps^2(J_c^2-J_T^2)}.
\eeq
The expression~(\ref{jt2}) for $J_T^2$ as a function of $\eps$, $\a$ and $\del$
has been used to derive the equality on the extreme right.
The $1/t$ relaxation law~(\ref{ctime})
holds whenever the initial value $J(0)$,
is on the attractive side of the critical point,
i.e., $-1<J(0)<J_c^\L$ in the left critical case (where $J_c^\L<J_T$ and so $A_c^\L<0$),
or $J_c^\R<J(0)<1$ in the right critical one (where $J_c^\R>J_T$ and so $A_c^\R>0$).
In the opposite regimes
($J_c^\L<J(0)<1$ in the left critical case or $-1<J(0)<J_c^\R$ in the right
critical case), one finds exponential relaxation to $J_2$ and $J_1$ respectively.

\subsection{Tricritical dynamics}

\begin{figure}[!ht]
\begin{center}
\includegraphics[angle=-90,width=.65\linewidth]{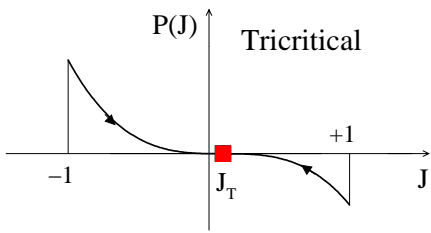}
\caption{\small
(Color online)
Tricritical dynamical behaviour ($J_1=J_2=J_3=J_T$).}
\label{tricritical}
\end{center}
\end{figure}

When all three fixed points merge at some $J_T$,
the dynamical system~(\ref{djdt}) exhibits a pitchfork bifurcation~\cite{bif}.
In physical terms, this corresponds to tricritical behaviour.
We have then
\beq
P(J_T)=P'(J_T)=P''(J_T)=0,
\label{triple}
\eeq
so that the tricritical synaptic strength $J_T$
is a triple zero of the rate function $P(J)$ (see Figure~\ref{tricritical}).

The mean synaptic strength again exhibits a universal power-law relaxation
(even slower than at the critical points) to its tricritical value $J_T$:
\beq
J(t)-J_T\approx\pm\frac{B_T}{\sqrt{t}}.
\label{ttime}
\eeq
This $1/\sqrt{t}$ relaxation law holds irrespective of the initial
value $J(0)$,
with $\pm$ denoting the sign of the initial difference $J(0)-J_T$, whereas
\beq
B_T=\sqrt{-\frac{3}{P'''(J_T)}}=\frac{1}{\sqrt{8\del\eps^2J_T}}.
\eeq
As $J_T$ is always positive,
the expression on the right-hand side is always well-defined.
We have actually $J_T>\frat{1}{3}$ (see~(\ref{tiers})),
and so the tricritical point too is always strengthening.

To sum up, the non-critical fixed points of Regimes~I or II
are characterised by exponential relaxation;
the corresponding relaxation times, whether long or short, are always finite.
Anywhere along the critical manifold, on the other hand,
one observes a universal power-law relaxation in $1/t$
of the mean synaptic strength.
An even slower power-law relaxation in $1/\sqrt{t}$ holds at the tricritical point.
These two cases correspond to an infinite relaxation time.

\section{Dependence on parameters and phase diagram}
\label{diagram}

We have so far described the various dynamical regi\-mes
characterising the evolution of the mean synaptic strength $J(t)$
according to the mean-field dynamical equation~(\ref{djdt}).
Here, we describe the regions of parameter space where these will be found.

\subsection{Dependence on the spontaneous rates}

It is worth examining first the phase diagram of the model
in the $\o$--$\O$ plane of the spontaneous rates,
for fixed values of $\eps$, $\a$ and $\del$.
This plane is also the arena where input signals are expressed (see Section~\ref{landf}).

The criticality conditions~(\ref{double}) allow us to express the critical
values
of the spontaneous rates in terms of $J_c$ as
\beqa
\o_c\!\!&=&\!\!\frat12(-3p_4J_c^4+4p_4J_c^3-p_2J_c^2+2p_2J_c-\a-\del),
\nonumber\\
\O_c\!\!&=&\!\!\frat12(3p_4J_c^4+4p_4J_c^3+p_2J_c^2+2p_2J_c-\a+\del).
\label{wcwwc}
\eeqa

At the tricritical point, the third equality of~(\ref{triple})
determines the value of $J_T$ as:
\beq
J_T^2=-\frac{p_2}{6p_4}=\frac{1}{6}\left(\frac{\a+\del}{\del}+\frac{1}{\eps^2}\right).
\label{jt2}
\eeq
From now on, $J_T$ will denote the {\it (positive)} square root of this expression.

The expressions~(\ref{wcwwc}) of the critical rates imply
\beqa
\frad{\partial\o_c}{\partial J_c}\!\!&=&\!\!-6\del\eps^2(1-J_c)(J_c^2-J_T^2),
\nonumber\\
\frad{\partial\O_c}{\partial J_c}\!\!&=&\!\!-6\del\eps^2(1+J_c)(J_c^2-J_T^2).
\eeqa
Viewed as functions of $J_c$, both critical rates are therefore simultaneously stationary
(i.e., either maximal or minimal) for $J_c=\pm J_T$.
The only way for the model to have a physical tricritical point,
where both spontaneous rates $\o_T$ and $\O_T$ are positive,
is to have $\del>0$, i.e., $\g>\b$, and $J=J_T>0$.
The spontaneous rates are then {\it maximal} at this tricritical point.

\begin{figure}[!ht]
\begin{center}
\includegraphics[angle=-90,width=.65\linewidth]{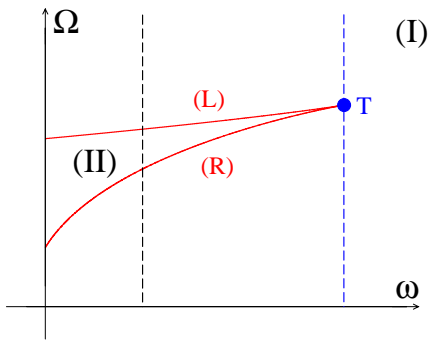}
\caption{\small
(Color online)
Schematic phase diagram in the $\o$--$\O$ plane.
T: tricritical point.
(L) and (R): left and right critical branches.
(I) and (II): Regimes~I and II.
Vertical dashed lines: cuts along which fixed points and relaxation times
will be investigated in Section~\ref{reltimes}
and plotted in Figures~\ref{cjtau} and~\ref{tjtau}.}
\label{www}
\end{center}
\end{figure}

Figure~\ref{www} shows a schematic phase diagram in the $\o$--$\O$ plane.
The horn-shaped curve is the critical manifold
ending in a cusp singularity at the tricritical point T.
The upper branch (L) corresponds to left critical dynamics,
while the lower one (R) corresponds to right critical dynamics.
The bounded region inside the critical curve corresponds to Regime~II,
while the complementary region corresponds to Regime~I.

\subsection{Dependence on the other parameters}

Let us now examine the phase diagram of the model
as a function of the remaining parameters $\eps$, $\a$ and $\del$.
The two latter rates only enter through their ratio,
which suggests the definition of the dimensionless quantity
\beq
g=\frac{\del}{\a+\del}.
\eeq
The tricritical values $\O_T$ and $\o_T$
of the spontaneous rates $\O$ and $\o$ turn out to obey $\O_T>\o_T$.
The condition for having a physical critical manifold resembling Figure~\ref{www},
culminating in a physical tricritical point, is thus $\o_T>0$.

\begin{figure}[!ht]
\begin{center}
\includegraphics[angle=-90,width=.65\linewidth]{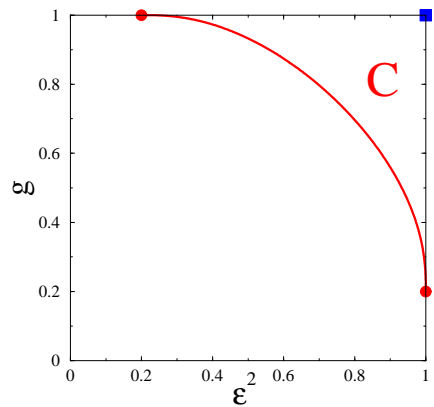}
\caption{\small
(Color online)
Phase diagram in the $\eps^2$--$g$ parameter space.
The model has a physical critical manifold in the region marked with C.
(Red) curve ending at filled circles: phase boundary (see~(\ref{red}))
with endpoints ($\eps^2=\frat{1}{5}$, $g=1$) and ($\eps^2=1$, $g=\frat{1}{5}$).
(Blue) filled square: extremal model ($\eps^2=g=1$).}
\label{e2g}
\end{center}
\end{figure}

Figure~\ref{e2g} shows the phase diagram of the model
in the $\eps^2$--$g$ parameter space (the unit square).
The (red) curve with equation
\beq
128\eps^2g(\eps^2+g)^3=3(\eps^4+14\eps^2g+g^2)^2
\label{red}
\eeq
is the phase boundary, corresponding to $\o_T=0$.
This curve exhibits an unexpected symmetry under the exchange of $\eps^2$ and $g$.
The endpoints, shown as red symbols,
have coordinates ($\eps^2=\frat{1}{5}$, $g=1$) and ($\eps^2=1$, $g=\frat{1}{5}$).

The model exhibits critical behaviour
only when the parameters $\eps^2$ and $g$ lie above the red curve,
i.e., in the rather small region marked C.
These rather stringent limitations on critical behaviour
suggest that the associated infinitely large relaxation time is only rarely observed:
most of the phase diagram is dominated by exponential relaxation.
This is consistent with the fact that one would expect infinitely large relaxation times
(with their possible association with long-term memory, see Section~\ref{landf})
to be associated only with rare events.

The upper right corner of Figure~\ref{e2g}, shown as a (blue) square,
corresponds to the {\it extremal model} where $\eps^2=1$
(the slope of the response function is maximal)
and $g=1$, i.e., $\a=0$ (absence of Hebbian learning).
The tricritical synaptic strength assumes the value
\beq
J_T=\frac{1}{\sqrt{3}}\approx0.57735.
\label{jt}
\eeq
In reduced time units where $\del=1$, the corresponding spontaneous rates read
\beqa
\o_T\!\!&=&\!\!\frat{2}{9}(2\sqrt{3}-3)\approx0.10313,
\nonumber\\
\O_T\!\!&=&\!\!\frat{2}{9}(2\sqrt{3}+3)\approx1.43646.
\label{ot}
\eeqa

The critical manifold is the largest possible in this extremal model.
With the notation of Figure~\ref{www},
the left (L) branch corresponds to $\frat{1}{3}<J_c^\L<J_T$,
while the right (R) branch corresponds to $J_T<J_c^\R<1$.
It can be checked that the range of possible values of $J_c$
is always smaller for generic parameter values in region C of Figure~\ref{e2g},
than in this extremal model.
In particular, we always have
\beq
J_c>\frat{1}{3}.
\label{tiers}
\eeq

\subsection{Relaxation times}
\label{reltimes}

In this section we illustrate the behaviour of the attractive fixed points
of the mean-field dynamical equation~(\ref{djdt}),
and of the corresponding relaxation times.
The main emphasis is on the divergent behaviour of the relaxation times
when the critical manifold or the tricritical point is approached.
The subsequent numbers and figures correspond to the extremal model ($\eps^2=g=1$).
This choice is only made for convenience;
any point within region C of Figure~\ref{e2g} would lead to a similar picture.
Finally, we work in reduced time units ($\del=1$).

\subsubsection{Critical behaviour}

In order to investigate the effect of the critical manifold,
we fix a value $\o=0.03$ and move along the left (black) vertical line of Figure~\ref{www}
by varying $\O$.
By so doing, we cross the left critical branch (L) at $\O_c^\L$
and the right critical branch (R) at $\O_c^\R$.

\begin{figure}[!ht]
\begin{center}
\includegraphics[angle=-90,width=.65\linewidth]{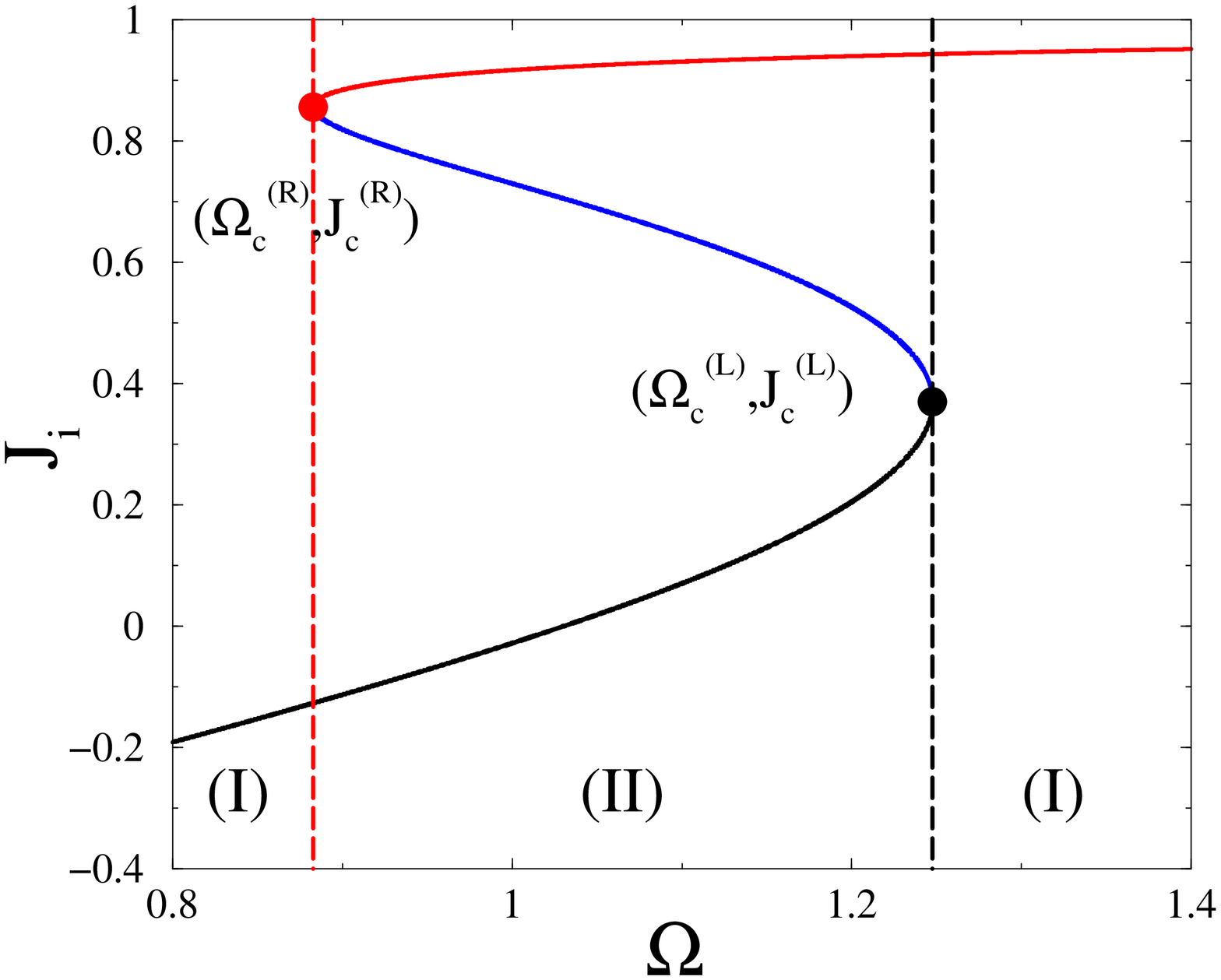}

\includegraphics[angle=-90,width=.65\linewidth]{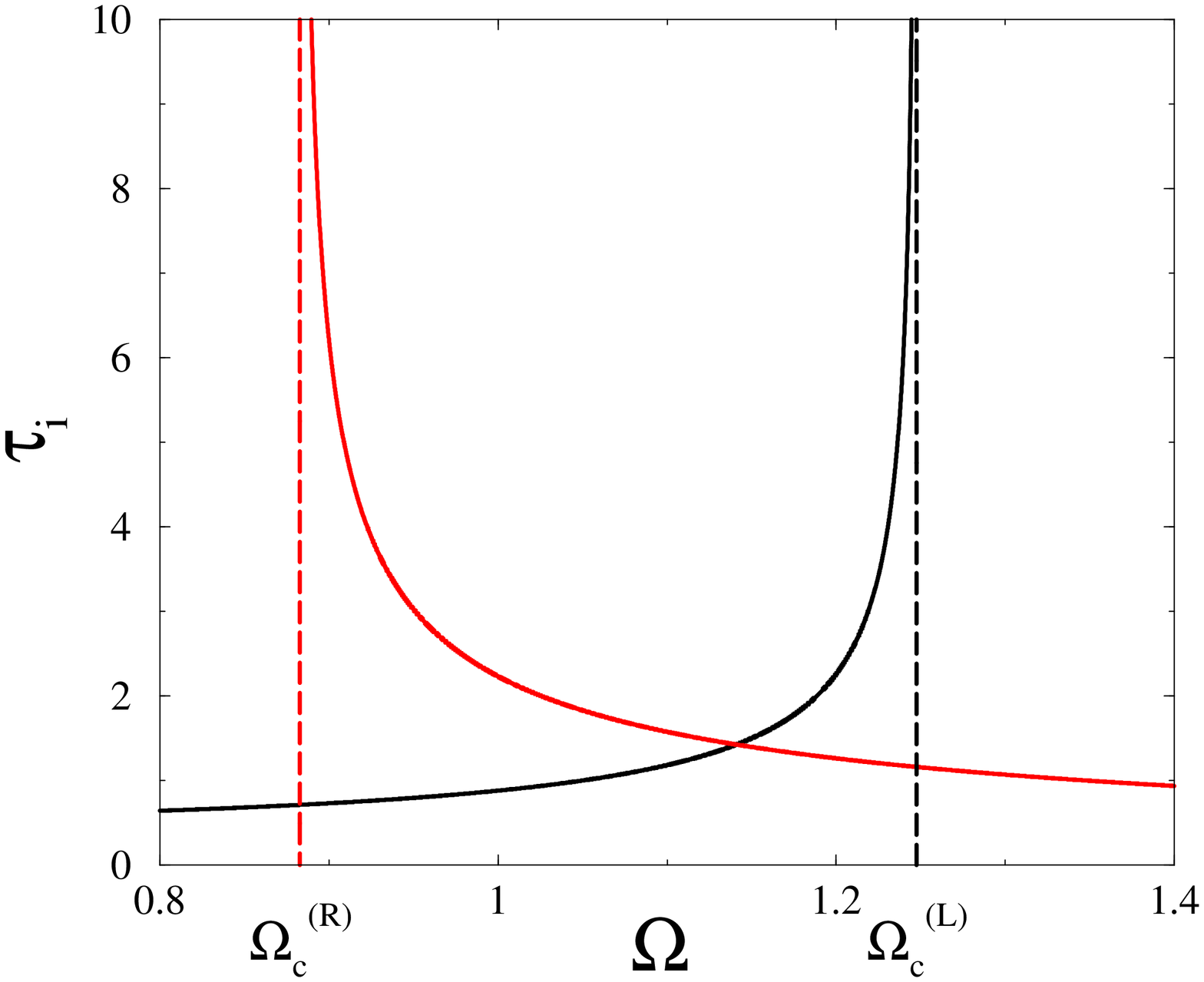}
\caption{\small
(Color online)
Top: Fixed points $J_i$ against potentiating rate $\O$ in reduced units,
for the extremal model with $\o=0.03$.
Bottom (black) and top (red) curves: attractive fixed points.
Intermediate (blue) curve: repulsive fixed point.
Right (black) and left (red) filled symbols have respective coordinates
$(\O_c^\L\approx1.24768,\,J_c^\L\approx0.37013)$
and $(\O_c^\R\approx0.88270,\,J_c^\R\approx0.85650)$.
Bottom: relaxation times associated with the attractive fixed points.
Vertical lines at $\O=\O_c^\L$ and $\O=\O_c^\R$ demarcate regimes I and II
and locate the divergences~(\ref{c1}),~(\ref{c2}).}
\label{cjtau}
\end{center}
\end{figure}

Figure~\ref{cjtau} shows the fixed points (top)
and the corresponding relaxation times (bottom)
against the potentiating rate $\O$.
In Regime~II, i.e., for $\O_c^\R<\O<\O_c^\L$,
the figure shows the two attractive fixed points, $J_1$ (lower (black) branch)
and $J_2$ (upper (red) branch)
and the two associated relaxation times, $\tau_1$ and $\tau_2$.
The intermediate repulsive fixed point $J_3$ (blue) is also shown for completeness.
One relaxation time diverges as each branch of the critical manifold is reached.
As $\O\to\O_c^\L$, where $J_1$ and $J_3$ merge at $J_c^\L$
(see Figure~\ref{critical}, top), we have
\beq
J_c^\L-J_1\sim\bigl(\O_c^\L-\O\bigr)^{1/2},\quad
\tau_1\sim\bigl(\O_c^\L-\O\bigr)^{-1/2}.
\label{c1}
\eeq
Similarly, as $\O\to\O_c^\R$, where $J_2$ and $J_3$ merge at $J_c^\R$
(see Figure~\ref{critical}, bottom), we have
\beq
J_2-J_c^\R\sim\bigl(\O-\O_c^\R\bigr)^{1/2},\quad
\tau_2\sim\bigl(\O-\O_c^\R\bigr)^{-1/2}.
\label{c2}
\eeq
Outside the interval $\O_c^\R\le\O\le\O_c^\L$, we are in Regime~I.
There is one single fixed point $J_0$ (see Figure~\ref{regimes}, top).
This fixed point appears as a continuation of $J_1$ for $\O<\O_c^\R$,
and as a continuation of $J_2$ for $\O>\O_c^\L$.

\subsubsection{Tricritical behaviour}

\begin{figure}[!ht]
\begin{center}
\includegraphics[angle=-90,width=.65\linewidth]{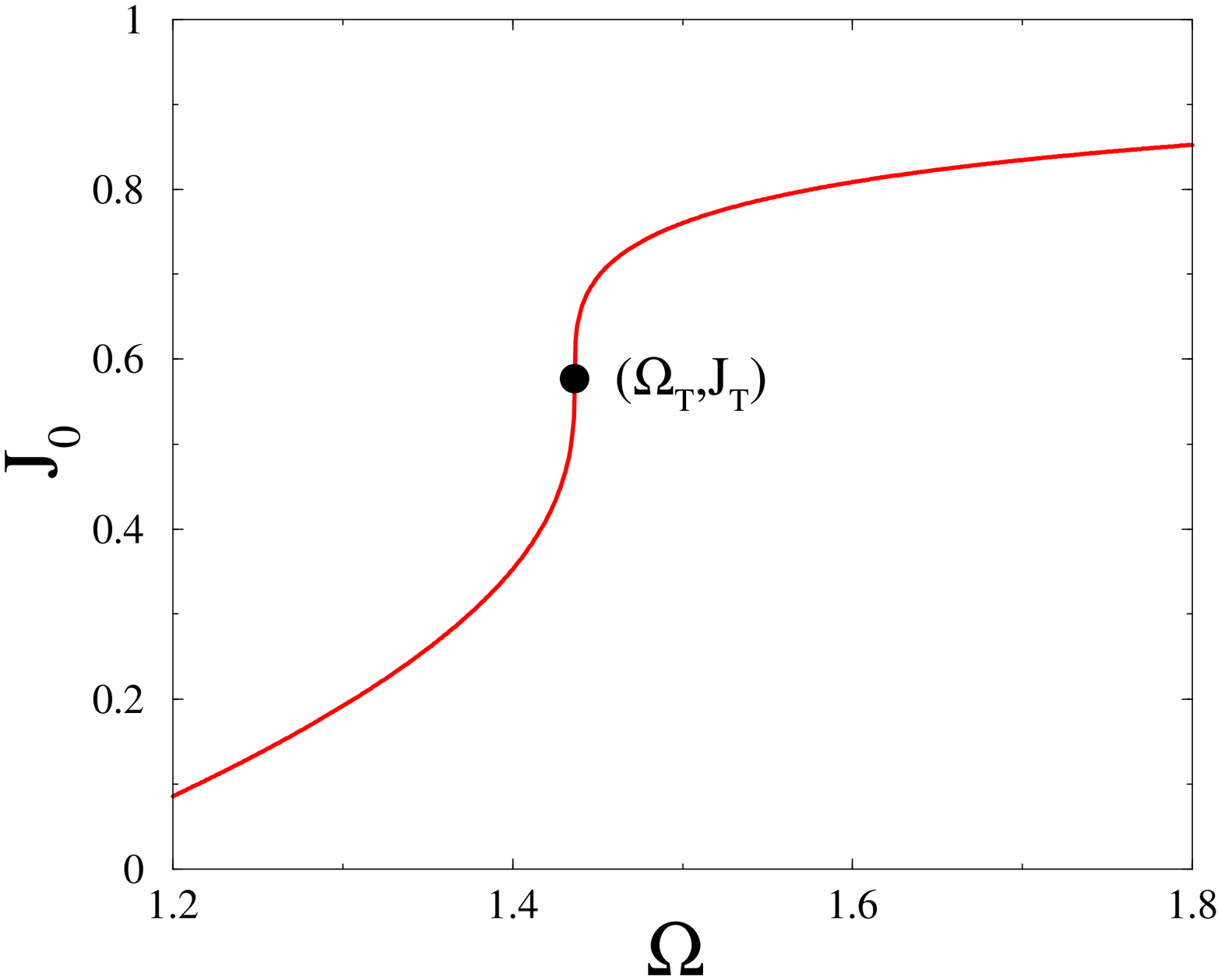}

\includegraphics[angle=-90,width=.65\linewidth]{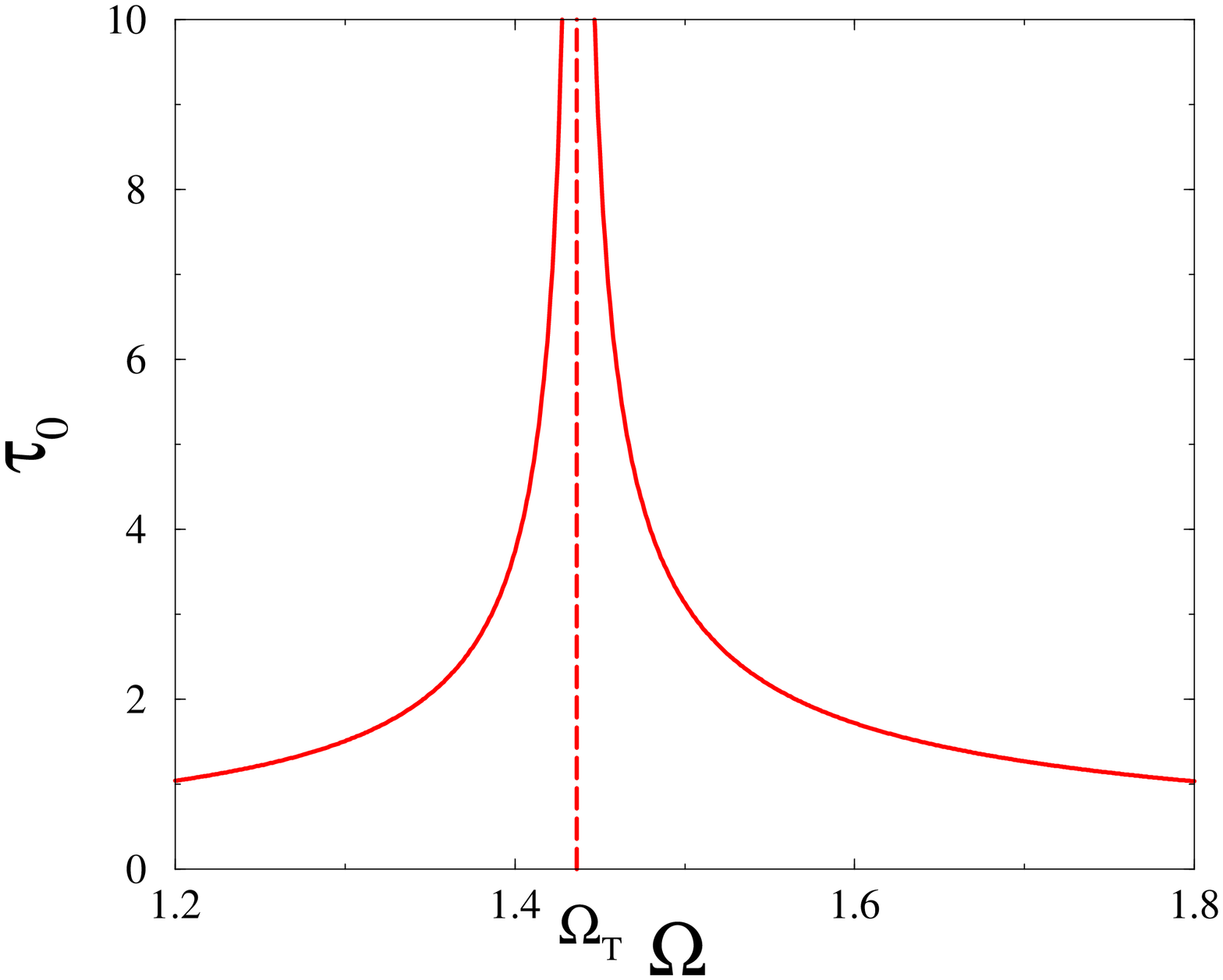}
\caption{\small
(Color online)
Top: attractive fixed point $J_0$ against potentiating rate $\O$ in reduced
units,
for the extremal model with $\o=\o_T$.
The coordinates $(\O_T,J_T)$ of the tricritical point (filled symbol)
are given by~(\ref{jt}),~(\ref{ot}).
Bottom: associated relaxation time $\tau_0$.
The vertical line at $\O=\O_T$ locates the divergence~(\ref{tc}).}
\label{tjtau}
\end{center}
\end{figure}

Now, in order to investigate the effect of the tricritical point,
we set $\o =\o_T$ (see~(\ref{ot})) and vary $\O$,
which traces the right (blue) vertical line of Figure~\ref{www}.
Consequently, we are always in Regime~I,
with its single fixed point~$J_0$ and corresponding relaxation time $\tau_0$.
These quantities are plotted in Figure~\ref{tjtau}.
As the tricritical point is approached ($\O\to\O_T$),
we have the power laws
\beq
\abs{J_0-J_T}\sim\abs{\O-\O_T}^{1/3},\quad
\tau_0\sim\abs{\O-\O_T}^{-2/3}.
\label{tc}
\eeq

\subsubsection{Summary}

We now summarise the content of the above paragraphs.
In the critical regime, the power laws~(\ref{c1}),~(\ref{c2}) are
characteristic of a saddle-node bifurcation.
They together conspire to hint at the slow relaxation in $1/t$ (see~(\ref{ctime}))
of the mean synaptic strength at a critical point.
Similarly, in the tricritical regime,
the relaxation in $1/\sqrt{t}$ (see~(\ref{ttime})) at the tricritical point
results from combining the power laws~(\ref{tc}),
which are characteristic of a pitchfork bifurcation.
The first thing worth
remarking is that the divergence of the tricritical relaxation time is
symmetric when $\O_T$ is approached from smaller and larger $\O$.
This is not the case for the critical regime, when the divergence
occurs at $\O_c^\L$ when approached from below
and at $\O_c^\R$ when approached from above.
Second, the faster growth of the relaxation time around the tricritical point causes
the overall slower relaxation of the mean synaptic strength,
with respect to the critical regime.

We end this section with a qualitative picture of our phase diagram
and its associated flows.
The choice of~$g$ and $\eps$ defines a specific model
within the region C of Figure~\ref{e2g}.
The behaviour of the latter as a function of $\O$ and $\o$
is illustrated in Figure~\ref{www}.
Here, $\o$ can be chosen such that, upon varying $\O$, the system is:

\begin{itemize}

\item fully confined to the noncritical Regime~I ($\o>\o_T$)
so that only exponential relaxation is possible.

\item constrained to reach the tricritical point ($\o=\o_T$).
Here, all initial configurations of synapses (ranging from totally weak to totally strong)
are attracted towards a unique tricritical point
which is strengthening ($J_T>0$).
The consequent power-law relaxation ($\sim 1/\sqrt{t}$) is at its slowest here.

\item free to explore the critical region ($\o<\o_T$).
For $\O<\O_c^\R$ and $\O>\O_c^\L$, the system is in Regime~I, and all
relaxation is exponential.
For $\O_c^\R<\O<\O_c^\L$, whether the critical point is reached or not
depends strongly on the initial synaptic configuration through $J(0)$.
While both fixed points are strengthening,
the important difference with the tricritical scenario
is that the associated power-law forgetting is faster ($\sim 1/t$) here.

\end{itemize}

The relaxation time for the mean synaptic strength
is likely to provide an upper bound for the relaxation time of specific
patterns stored in a distributed fashion across the network.
Consequently, as will be discussed in the next section,
the critical and tricritical situations
are those where power-law forgetting can be manifested.

\section{Learning and forgetting}
\label{landf}

\subsection{Global properties}

In this last section we address the learning and forgetting properties
of the model network.
We start with global features,
by submitting our model to an arbitrary time-dependent but spatially uniform input.
The latter is modelled by two deterministic signals $S(t)$ and $s(t)$,
which are respectively superposed on the spontaneous relaxation rates, according to
\beqa
\O(t)\!\!&=&\!\!\O+S(t),
\nonumber\\
\o(t)\!\!&=&\!\!\o+s(t).
\label{global}
\eeqa

The single collective degree of freedom of the model,
namely its mean synaptic strength $J(t)$,
evolves according to equation~(\ref{djdt}), where the rate function
\beqa
P(J;t)\!\!&=&\!\!p_4J^4+p_2J^2
\nonumber\\
\!\!&-&\!\!(\O+\o+\a+S(t)+s(t))J
\nonumber\\
\!\!&+&\!\!\O-\o-\del+S(t)-s(t)
\label{pjtime}
\eeqa
now bears an explicit time dependence.

Let us assume for definiteness
that the mean synaptic strength has relaxed to one of its fixed-point values $J$,
and that the signals $S(t)$ and $s(t)$
are non-zero only in a finite time window of duration $T$.
During the learning phase ($0<t<T$),
$J(t)$ will be displaced from the fixed-point value $J$,
characterising its {\it default state} in the absence of any input.
During the subsequent forgetting phase ($t>T$),
$J(t)$ will relax back to its default state.

If the default parameters are such that the system lies in the non-critical regime,
then both learning and forgetting will be exponentially fast.
Optimal trajectories can in principle be constructed,
as was done in~\cite{gmam1,gmam2} so that fast learning and slow(er) forgetting
are obtained, but globally, the memory manifested is always short-term.
If, however, the default state of the model is critical,
the application of generic input signals will take the system off it,
so that the learning mechanism will be characterised by a finite relaxation time.
Learning will thus be exponentially fast,
while forgetting (at the same global level)
will follow the power law~(\ref{ctime}) characteristic of the critical state.

\subsection{Local properties}

Associative memory is usually encoded in patterns
which are stored throughout neural networks in a distributed fashion,
i.e., as a non-uniform modulation of the synaptic weights $\s_{ij}$.
In order to explore the storage of memory in our model,
an arbitrary space and time-dependent input
(modelled as deterministic signals $S_{ij}(t)$ and $s_{ij}(t)$) would have to be
superposed on the spontaneous relaxation rates of every synapse $(ij)$:
\beqa
\O_{ij}(t)\!\!&=&\!\!\O+S_{ij}(t),
\nonumber\\
\o_{ij}(t)\!\!&=&\!\!\o+s_{ij}(t).
\label{local}
\eeqa

The resulting equations can only be investigated by means of extensive
numerical work in the general case.

We can however look analytically at the response of a {\it single} synapse, say $(kl)$,
when it is submitted to an input such as~(\ref{local}).
The mean synaptic strength $J(t)$ of the whole network will be unaffected
by such a localised perturbation, and thus continue to obey~(\ref{djdt}).
Let us again assume that the system has reached one of the fixed-point values $J$.
The stochastic dynamics of the selected synapse $(kl)$ can be shown,
along the lines of the derivation of~(\ref{p123}),
to be determined by the effective rates
\beqa
\O_\eff(t)\!\!&=&\!\!\O+\frat12\a(1+\eps^2J^2)
\nonumber\\
\!\!&+&\!\!\frat{1}{4}\b(1+J)(1-\eps^2J^2)+S_{kl}(t),
\nonumber\\
\o_\eff(t)\!\!&=&\!\!\o+\frat12\a(1-\eps^2J^2)
\nonumber\\
\!\!&+&\!\!\frat{1}{4}\g(1-J)(1-\eps^2J^2)+s_{kl}(t).
\label{wwweff}
\eeqa
The mean strength $j_{kl}(t)$ of the selected synapse therefore obeys
\beq
\frac{\d j_{kl}}{\d t}=\O_\eff(t)(1-j_{kl})-\o_\eff(t)(1+j_{kl}).
\eeq
In the forgetting phase ($t>T$), the input signals vanish,
so that the mean strength of the selected synapse relaxes to the fixed-point value $J$
of the mean synaptic strength of the whole network.
Interestingly, this relaxation is always exponential:
\beq
j_{kl}(t)-J\sim\e^{-t/\tau_\loc}.
\eeq
The corresponding local relaxation time is such that its reciprocal
is the sum of both rates~(\ref{wwweff}) in the absence of a signal, i.e.,
\beq
\frac{1}{\tau_\loc}=\O+\o+\a+\frat{1}{4}(\b(1+J)+\g(1-J))(1-\eps^2J^2).
\eeq

This relaxation time is likely to provide a lower bound for
time scales associated with short-term memory.

\subsection{A rich spectrum of time scales}

As indicated above, networks learn by assimilating space- and time-dependent patterns.
Realistic learning and forgetting protocols depend on the precise space-
and time- dependence of applied signals, as well as, of course, the default
parameters of the network.
In the previous two subsections, we have shown that global time scales associated
with the dynamics of the entire network can be either finite (exponential relaxation)
or infinite (power-law relaxation),
while the relaxation of a single synapse is always exponential.
The global and local time scales
provide estimates for the upper and lower bounds respectively
for learning and forgetting in realistic situations,
which can involve all possible time scales in between.
This generation of such a rich `dispersive' spectrum of time scales
from a simple model of a synaptic network has enabled us to unify
the hitherto somewhat separate~\cite{fusi,jstat} domains of modelling long-
and short-term memory.

We remark that the default parameters of the model correspond to the intrinsic
properties of the network: given this, it is rather fitting that the critical
manifold is rather small, and has to satisfy rather stringent conditions in order to exist.
In other words: while exponential forgetting is generic, one needs
to design networks rather carefully to get long-term memory storage.

\section{Discussion}
\label{disc}

Memories can be short- or long-lasting.
Our ability to store information depends
as much on our intrinsic neural structure, as well as, typically, the significance
of this information.
One of the most important features of the model presented here is that
it provides a natural framework for this separation of time scales in terms of
the default (intrinsic) state of the system and the nature of applied signals.
Short-term memories are forgotten exponentially fast,
with a whole spectrum of time scales determining how fast the forgetting actually is.
Power-law forgetting may hold for long-term memories corresponding to the
default state of the system being on a critical manifold;
there, the mean synaptic strength has a universal $1/t$ fall-off,
which gets even slower at the tricritical point,
where it is turned into a $1/\sqrt{t}$ behaviour.
Since the existing model
for long-term memory relies on possibly unrealistic auxiliary structures
such as internal states~\cite{fusi,jstat}, our model represents an
important conceptual advance in the field; it unifies the modelling of short-
and long-term memory,
which emerge naturally from collective synaptic dynamics, without
the need to invoke special architectures.

This framework is also an appropriate one to discuss optimal learning.
This occurs if the default state of a neural system lies on the
critical manifold, so that fast learning will occur for generic signals
(which will typically perturb the system to one of Regimes~I or~II),
while forgetting will be extremely slow.
The dynamics in Regime~II may exhibit yet other phenomena,
with the possibility of the degradation or improvement
of the same system as a result of strong enough applied signals.
The application of more complex protocols, i.e., time-dependent signals,
on the present model would reveal a very rich dynamical behaviour
with a whole panoply of possible scenarios.
As a rather extreme situation, one may think of complex learning protocols,
corresponding to signals cycling around the critical manifold
or the tricritical point in the $\o$--$\O$ plane,
and of the corresponding very unusual `ageing' behaviour.

We now make a few remarks on the design of our model.
The slow plasticity dynamics of synapses are
driven by competitive and cooperative interactions
consequent on the fast dynamics of firing neurons.
The model is analysed within a mean-field approximation,
in common with many physics-based approaches to neuroscience, ranging from earlier
work~\cite{ags1,ags2,ags3,ags4,nadal,parisi,landf1,landf2,landf3,landf4}
to more recent developments~\cite{sydney}.
Such a mean-field framework is of course appropriate
given the lack of knowledge of microscopic details at the neural or synaptic level.
Our model includes some ideas presented
in earlier work~\cite{gmam1} but improves on them by cleanly separating
the roles of neurons and synapses, as well as by introducing directedness.
At the microscopic level, the introduction of directedness is quite complex, since
for example, causality demands that synaptic updating occurs when a spike train from a
presynaptic neuron reaches a postsynaptic neuron; this could lead to rather
complicated rules reflecting such instantaneous, spike-time-dependent updates~\cite{song}.
At our level of description involving much longer time scales, however, such
spikes are averaged over and represented by a mean neural activity which
determines synaptic updates.

We should also add that the construction of
alternative synaptic update rules to incorporate
some aspects of this microscopic causality would, at least in the mean-field
perspective presented here, not make much difference.
In particular, introducing a non-linear constitutive function $g(J)$
would only result in more intricate equations
without changing any essential feature.
In other words, the critical phenomena exhibited by our model,
and the global features of its phase diagram, are robust to microscopic details.
The latter would of course regain their importance
if our model were to be investigated by means of computer simulations on large networks.

Also, we point out that the cooperative Hebbian mechanism turns out to be
almost entirely irrelevant to obtaining critical behaviour in our model.
On the other hand, a strong enough synaptic competitivity -- the most novel and
original feature of this work -- turns
out to be the crucial ingredient for the potential manifestation
of both short- and long-term memory in our model network.
Since the human brain is thought of as being a vast and complex
synaptic network which also has this remarkable ability to store memories across
a rich spectrum of time scales, our work underlines the current view that
mechanisms of synaptic competitivity are of critical importance in
neuroscience~\cite{avy}.

Finally, we make a few remarks on possible experiments to test our model.
There has been a great
deal of research into the idea that sleep consolidates short-term memories into
long-term ones~\cite{dborn,timofeev,yang}; experiments on rats
suggest that this transformation occurs via a hippocampus-cortical memory
transfer~\cite{sidarta}.
In the context of our model,
this would suggest that competitive mechanisms predominate in the cortex rather
than the hippocampus; this idea provides a testable prediction for experiments in vivo.
A possible in vitro experiment could involve the adaptation of high-resolution
measurements of cultured cortical slices~\cite{friedman}, which have been very
successful in probing neuronal avalanches, to the present situation: since
these methods are able to distinguish clearly between exponential and power-law
signatures in neuronal avalanches, one might reasonably hope that they would be
able to do the same in the context of stored synaptic memories.

\begin{acknowledgments}

AM warmly acknowledges the hospitality of the Institut de Physique Th\'eorique,
where most of this work was conceived and carried out.
She also thanks Konstantin Klemm for useful discussions on models of directed synapses.

\end{acknowledgments}

\bibliography{mybibfile.bib}

\end{document}